\documentclass[a4paper, 10pt, conference]{ieeeconf}      

\IEEEoverridecommandlockouts                              

\usepackage{graphics} 
\usepackage{epsfig} 
\usepackage{url}

\title{\LARGE \bf
Representing, Simulating and Analysing\\
Ho Chi Minh City Tsunami Plan\\
by Means of Process Models
}

\author{Nguyen Tuan Thanh LE$^{1}$, Chihab HANACHI$^{2}$, Serge STINCKWICH$^{3}$, and Tuong Vinh HO$^{4}$
\thanks{$^{1}$Nguyen Tuan Thanh Le is with Toulouse 1 University and member of the IRIT Laboratory (SMAC team), France}
\thanks{$^{2}$Chihab Hanachi is with Toulouse 1 University and member of the IRIT Laboratory (SMAC Team), France
       {\tt\small hanachi@univ-tlse1.fr}}%
\thanks{$^{3}$Serge Stinckwich is with UCBN \& UMI 209 UMMISCO (IRD/UPMC), France
       {\tt\small serge.stinckwich@ird.fr}}%
\thanks{$^{4}$Tuong Vinh Ho is with MSI/IFI/Vietnam National University \& UMI 209 UMMISCO (IRD/UPMC), Vietnam
       {\tt\small ho.tuong.vinh@auf.org}}%
}

\begin{document}

\maketitle
\thispagestyle{empty}
\pagestyle{empty}

\begin{abstract}
This paper considers the textual plan (guidelines) proposed by People's Committee of Ho Chi Minh City (Vietnam) to manage earthquake and tsunami, and try to represent it in a more formal way, in order to provide means to simulate, analyse and adapt it. We first present a state of the art about coordination models for disaster management with a focus on process oriented approaches. We give an overview of the different dimensions of the textual tsunami plan of Ho Chi Minh City and then the graphical representation of its process with BPMN (Business Process Model and Notation). We finally show how to exploit this process with workflow tools to simulate ({\sc YAWL} tool) and analyse it ({\sc ProM} tool).
\end{abstract}

\section{INTRODUCTION}

	
Vietnam, giving its particular geographic location in the South-East Asian region, could get hit by natural disasters like tsunami or earthquake. The risk of tsunami in Vietnam is not very large, but exists, and it is necessary to prepare for the disaster (\cite{Ca:2008:tsunami:vietnam}). Plans to manage that kind of crisis are most of the time available in a textual format defining the actors, their roles and their coordination in the different steps of crisis life-cycle: \emph{mitigation}, \emph{preparedness}, \emph{response} and \emph{recovery}. While plans in a printed document format are easy to manipulate by stakeholders taken individually, they do not provide direct means to be analysed, simulated, adapted, improved and may have various different interpretations, so difficult to manage in real time and in a distributed setting. Besides, organizations in charge of crisis management require team collaboration tools based on information and communication technologies that facilitate the distribution of control, and information exchanges and gathering. Most of the time, tools still remain limited to basic ones such as whiteboard, radio, telephone, word processor or spreasheets, that do not support the whole plan, viewed as an explicit, shared and dynamic coordination workspace, providing an overview of the crisis state and evolution.

Giving these observations, it becomes useful to model these textual plans to have an accurate representation of them, to reduce ambiguity, to support coordination between stakeholders and ease an efficient control  and resolution of the crisis.

In this paper, we consider a concrete case study which is a textual document defined by Ho Chi Minh City and that specifies the plan to be undertaken in order to deal with dedicated events such as tsunami and/or earthquake.

We consider the Ho Chi Minh City Crisis textual plan and then we discuss and provide coordination models to support the corresponding crisis management, limited to a tsunami in this paper. Coordination is considered here according to a multi-agent view (\cite{Smith:2011:OC:1964041.1964044}), i.e a mean to support the interdependence between actors, the achievement of common goals, and the sharing of resources and competencies. Several possible approaches/techniques have been identified to support coordination: \emph{organizational structuring}, \emph{contracting}, \emph{negotiation} and \emph{planning}.

In our work, we will use a process oriented-approach (\cite{Van:2003:business:survey}), which represents plans in an organizational context. Indeed, as shown in (\cite{Sell:2009:workflow:emergency}), plans may be abstracted as (business) processes and therefore implemented as workflows. Processes are defined as a set of coordinated tasks, each task corresponding to a plan measure. A process model also includes an organizational model that defines the actors and resources, their roles, the task allocation policy and the actors relationships in terms of hierarchy and delegation. Following a process approach, it is possible to support the whole life cycle of a plan: \textit{description}, \textit{simulation}, \textit{analysis}, \textit{execution}, \textit{monitoring} and \textit{mining}.

In a nutshell, the problem being addressed in this paper is: how to represent textual plans by means of processes to ease their analysis and engineering, and in particular the Ho Chi Minh City Tsunami Plan.

Our contribution consists in modeling Ho Chi Minh City crisis textual plan according to a process approach, and in showing how it could be used to improve tsunami management mostly in the  Mitigation, Preparedness  and Recovery phases of a disaster management life-cycle. This work is a first step towards a process aware information system for crisis management. 

In the following section, we first present a state of the art about coordination models for disaster management with a focus on process oriented approaches. In section three, we summarize the Ho Chi Minh City tsunami management textual plan. In section four, we present a possible process model of this textual plan using the BPMN notation. Then we discuss the advantage of such a model, and notably how such a model can be simulate or executed with the YAWL workflow system, analysed, or mined with the PROM tool. Finally, a conclusion is provided with some directions for future work.

\section{Coordination models for\\ disaster management}
Coordination is recognized as a main issue in crisis management (\cite{Chen:2008:coordination:emergency}, \cite{Hanachi:2012:CTR:2360026.2360677}). Indeed, coordination is required to ensure the coherent behavior of the cooperative and distributed actors involved in the crisis resolution. Coordination in a multi-agent world, such as the one met in crisis management, has been deeply studied in the literature (\cite{Smith:2011:OC:1964041.1964044}, \cite{Omicini:2001:coordination:internetagents}). \cite{Smith:2011:OC:1964041.1964044} provides an ontology of coordination that provides a model structuring all the theoretical concepts (\cite{Malone:1994:coordination:interdisciplinary}) involved in coordination: actors and their relationships, resources, activities and their interdependence (negative and positive coordination), coordination rules.  Following the standard format usually applied in coordination theory (\cite{Busi:2001:coordination:models}), a coordination model may be viewed as a triplet (E, M, R) corresponding respectively to (i) the set of coordinated entities (agents' activities), (ii) the media or infrastructure supporting the coordination (map, blackboard, P2P channel, ...) and iii) the set of coordination rules governing the coordination.

\begin{figure}[h]
	\centering
    \includegraphics[scale=0.4]{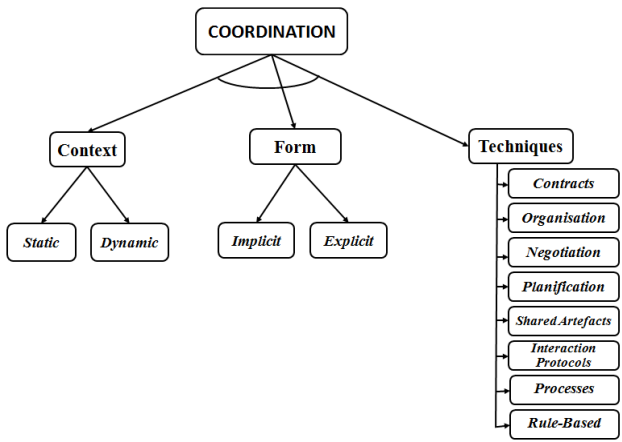}
    \caption{Coordination model parameters involved in a choice of a technique}
    \label{CoordinationModelParameters}
\end{figure}

Rules may be abstracted by coordination techniques such as contracts, organization, negotiation, planning, shared artifacts, interaction protocols, event-based rules or processes. As shown in Fig.~\ref{CoordinationModelParameters}, a coordination model may combine several techniques according to the nature of the universe considered (dynamic or static) and the form of rules chosen (explicit or implicit). While shared artifacts, such as maps, do not constraint the coordination, protocols are stricter, and event-based rules by their declarative style are between both. Processes can be considered as a combination of plans within organizational structure. The advantage of process-oriented coordination is to provide visibility on the whole crisis evolution (\cite{Franke:2013:coordinatio:framework:dynamic}): past, present and future activities and their relationships. Knowing by whom and how actions may be performed also eases accountability management and awareness. Whatever the (combination) of coordination model(s) followed, in a highly dynamic, open, unstable environment, such as the one met in crisis context, coordination should be adaptive and visible for all the partners while preserving the essential part of their autonomy. \cite{Schonenberg:2008:taxonomy} provides an overview of techniques to improve the flexibility and adaptation of processes. 

In this paper, we consider a process-oriented coordination that seems to us the most appropriate technique to represent the Ho Chi Minh City disaster management plan. We will consider the opportunities offered by such an approach in the context of BPM (Business Process Management) framework (\cite{Van:2003:business:survey}) which provides techniques, tools and notations to engineer processes: design, analyse, simulation, execution through a workflow system, monitoring, adaptation and process mining. Regarding coordination models aiming at representing plans, most of the works have considered event-based, artifact-based and/or process-aware models. \cite{Franke:2011:coordination:conflict} proposes a very detailed review of process management systems addressing a disaster response scenario. \cite{Ma:2012:research:management:disaster} proposes a coordination model of emergency response operations via an event-based awareness mechanism. Activities are represented as shared plans and a plan is updated each time an activity state evolves. Users subscribe to events (meaningful states transitions of activity, goal or resource) and are notified as soon as an event interesting to them occurs. \cite{Franke:2013:coordinatio:framework:dynamic} proposes an inter-organizational approach without the definition of a central process but by allowing organizations to share selected activities and by providing means for detecting and handling conflicts when the state of an activity changes. Each activity has a life-cycle similar to a micro-process. In the previous works, processes are in somehow implicit and neither considered as the main component of the coordination nor engineer during their whole life-cycle as we do. The most advanced works is the one developed in the Workpad project (\cite{Catarci:2011:workpad:process:disaster}) that shows the benefits to build a Process-aware Information System for emergency management, and proposes an architecture and a system to support the execution of emergency management processes. Rescue operators are supposed to be equipped with PDA and their work is orchestrated by a workflow system called ROME4EY hosted on the team leader’s device. This work focusses on the design and execution phases of process life-cycle and aims at supporting team works and not the global disaster plan as we consider. \cite{Sell:2009:workflow:emergency} shows how to use a workflow management system to manage emergency plans. This work is closer to us but does not address the whole process life cycle, focusses on resource management and delegation functionalities, and does not deal with a specific plan as we do with the Ho Chi Minh City one.

\section{The Ho Chi Minh City Tsunami Management Plan}


The Ho Chi Minh City tsunami response plan (\cite{PCHCMCity:2012:decision:3558QDUBND}) is a textual document\footnote{\url{http://www.phongchonglutbaotphcm.gov.vn/?id=51&cid=4395}} (44 pages) defined by People's Committee of Ho Chi Minh City in 2012. It specifies how to manage natural crisis situations (Earthquake level VI with no tsunami, Earthquake level VII with no tsunami, offshore earthquake with tsunami warning, ...) and overcome its consequences. 


It defines the roles and the tasks of each stakeholder, the mobilized resources and the organization of communication. This plan ensures the unified direction, the tight coordination between organizations in order to meet the response time requirement. It is also the foundation for the stakeholders in developing their own action plan according to their assigned duties; in proposing investment budget, building physical facilities, buying facilities and equipments for human resources development, training techniques to serve search and rescue tasks; and in deploying propaganda tasks, raising people's awareness to respond actively and effectively to minimize damage caused by the tsunami/earthquake.


It covers the following phases of the Disaster Management Life Cycle:
\begin{itemize}
\item \emph{Mitigation phase}. It is called in the Ho Chi Minh City Plan "the Propagation Mission". It corresponds to a long term preparation of both the crisis stakeholders and the citizens. It provides recommendations for organizing special classes for maintaining the stakeholders' knowledge about tsunamis/earthquakes. Also, regular propagations are planned to provide regularly information to citizens by mass media in order to educate them and help them to well-behave and avoid panics when a crisis occurs.
\item \emph{Preparedness phase}. It is called the Rehearsal Mission and it corresponds to real exercises made to improve the coordination of the stakeholders and to learn how to use the equipments fluently.
\item \emph{Response phase}. It is called the Period of Response and Search-Rescue. It gives the plan to be followed by the stakeholders: theirs tasks and how their have to coordinate (exchanges of information, and relation order between their tasks). It distinguishes four different scenarios (situations) which differs from their gravity. For each scenario, a precise plan is given.
\item \emph{Recovery phase}. It is called the Period of Overcoming the Consequences and Environmental Sanitation. It proposes means to recover a normal situation from transport, environment, health and public order points of view. It also provides recommendations on how to evaluate and learn from experience in order to adjust the plan.
\end{itemize}

In our work, we focus mainly on the Response phase in case of a tsunami. In response to a tsunami, local administrations, authorities, non-governmental organizations and inhabitants have to coordinate tightly.

We have identified in this plan 33 actors and several tasks for each actor. For legibility reason, we have gathered some actors with similar responsibility and mission in a more abstract actor. The actor \emph{Local Administration} represents the actually four following actors: 1) Committee for Flood-Storm Prevention and Search-Rescue of HCM City, 2) People's Committee of Districts, Communes and Towns, 3) Chairman of People's Committee of Districts, Communes and Towns, 4) Command Center of program against flooding of HCM City. The actor \emph{Communication unit} gathers three actors: 1) Department of Information and Communication, 2) Television station of HCM City, 3) Radio Voice of HCM City People. The actor \emph{Military} represents three actors: 1) High command of HCM City, 2) Border Guard High Command of HCM City, 3) Border Guard Forces. The actor \emph{Police} substitutes for two actors: 1) HCM City Police, 2) Department of Police in charge of Fire Prevention and Firefighting of HCM City. The actor \emph{Health and Red Cross organizations} gathers three actors: 1) Department of Health, 2) Center for Preventive Medical, 3) Red Cross of the City. The actor \emph{Local civil defense forces} represents two actors: 1) Local Civil Defense Forces, 2) Young Volunteers Force of HCM City.

All the actors considered in our model and theirs tasks are given in TABLE I.

\subsection{Actors and Tasks} 
During the response and search-rescue period, the 7 following actors are involved:

\begin{itemize}
\item A1: Institute of Geophysics (Vietnam Academy of Science and Technology)
\item A2: Local administration
\item A3: Military
\item A4: Police
\item A5: Local civil defense forces
\item A6: Communication unit
\item A7: Health and Red Cross organizations
\end{itemize}

Each actor have many tasks. We list in Table I only the main tasks for each actor. In the next subsection, we describe the process of a typical situation for responding to a tsunami.

\begin{table}[h]
\label{Tasks}
\begin{center}
\begin{tabular}{|l|l|}
	\hline
	\textbf{TASKS} & \textbf{ACTORS}\\
	\hline
	T1: Detect tsunami risk & A1: Institute of Geophysics\\
	\hline
	T2: Inform tsunami start & A1: Institute of Geophysics\\
	\hline
	T3: Receive tsunami start & A2: Local administration\\
	\hline
	T4: Fire the flares & A2: Local administration\\
	\hline
	T5: Notify functional units to start & A2: Local administration\\
	\hline
	T6: Inform by portable speakers & A5: Local civil defense forces\\
	\hline
	T7: Broadcast over the media & A6: Communication unit\\
	\hline
	T8: Evacuate people & A3: Military\\
	\hline
	T8': Evacuate people & A4: Police\\
	\hline
	T9: Transfer injuries to safe places & A3: Military\\
	\hline
	T9': Transfer injuries to safe places & A4: Police\\
	\hline
	T10: Inform, guide the fisherman & A3: Military\\
	\hline
	T11: Protect the property & A4: Police\\
	\hline
	T12: Perform the first aid & A7: Health \& Red Cross\\
	\hline
	T13: Call ambulance & A7: Health \& Red Cross\\
	\hline
	T14: Detect tsunami end & A1: Institute of Geophysics\\
	\hline
	T15: Inform tsunami end & A1: Institute of Geophysics\\
	\hline
	T16: Receive tsunami end & A2: Local administration\\
	\hline
	T17: Notify functional units to end & A2: Local administration\\
	\hline	
	T18: Identify damages & A3: Military\\
	\hline
	T18': Identify damages & A4: Police\\
	\hline
	T19: Search distress fishermen & A3: Military\\
	\hline
	T19': Search distress fishermen & A4: Police\\
	\hline
	T20: Close crisis response & A2: Local administration\\
	\hline
\end{tabular}
\end{center}
\caption{Actors and their tasks in tsunami response system}
\end{table}

\subsection{Processes}
When detecting the risk of a tsunami (\emph{T1}) affecting Ho Chi Minh City, the Institute of Geophysics (\emph{A1}) will inform (\emph{T2}) city's local administration (\emph{A2}) about the time, place and predictive level of the tsunami so that it can be prepared to respond. After receving this warning (\emph{T3}), the local administration (\emph{A2}) notifies (\emph{T5}) the tsunami risk to city's functional units such as: military (\emph{A3}), police (\emph{A4}), local civil defense forces (\emph{A5}), communication unit (\emph{A6}), health and Red Cross organization (\emph{A7}), ... so that they could initiate and coordinate their own activities. At the same time, the local administration (\emph{A2}) fires the flares (\emph{T4}) to warn the people. The local civil defense forces (\emph{A5}) use portable loudspeakers (\emph{T6}) walking along all city's streets and the residential areas to notify citizens in order that they should be evacuated to the safe places under the guidance of the military (\emph{A3}) or the police (\emph{A4}). Meanwhile, communication unit (\emph{A6}) broadcasts the warning message (\emph{T7}) on radio and television to inform people on the shore as well as on the vessels about the tsunami risk. The military (\emph{A3}) and the police (\emph{A4}) evacuate people (\emph{T8}) to the safe shelters. In addition, the military  transfers (\emph{T9})  injured people and distress fishermen to relief places and the police protects (\emph{T11}) the citizens' property. In parallel, the navy (\emph{A3}) whistles alarm, fires flares to inform and guide fishermen (\emph{T10}) to keep their boats in a safe location. The health unit and Red Cross organization (\emph{A7}) ensure the first aids (\emph{T12}) for injured people, for the serious case, they have to call the ambulance (\emph{T13}) to transport the injuries to the hospital. 

When the Institute of Geophysics (\emph{A1}) detects the tsunami has passed (\emph{T14}), it informs (\emph{T15}) the Ho Chi Minh City's local administration (\emph{A2}). After receiving this message (\emph{T16}), the city's local administration (\emph{A2}) announces it (\emph{T17}) to the functional units and citizens through the media. The military (\emph{A3}) and police (\emph{A4}) coordinate to search the distress fishermen (\emph{T18}) on the sea and identify the damages (\emph{T19}). Finally, the local administration (\emph{A2}) closes the tsunami response (\emph{T20}).

\section{Representing a textual plan by means of a workflow process}
We have now clearly identified the actors, the tasks and an informal process. What is the next step ? How can we use this information ? In our work, we choose to represent it as a workflow process. Workflow technology is relevant to represent crisis resolution plan for several reasons: 
\begin{itemize}
\item Workflow models capture emergency plans as demonstrated in \cite{Sell:2009:workflow:emergency}.
\item Once plans modelled as processes, they can be implemented, simulated, checked, analysed or executed through workflow management systems tools like YAWL. Mining tools like {\sc ProM}\footnote{\url{http://www.promtools.org/prom6/}} can also discover models from event logs and check if they are compliant with previous defined ones: often the real process can deviate from the prescribed one.
\item Workflow provides several description languages at different levels of abstraction. BPMN (Business Process Model and Notation) can be used for validation by end-users while Petri Nets serve theoretical analysis and simulation. Also tools like YAWL have their specific notations and provide operational models.
\end{itemize}

\begin{figure}[h]
	\centering
    \includegraphics[scale=0.3]{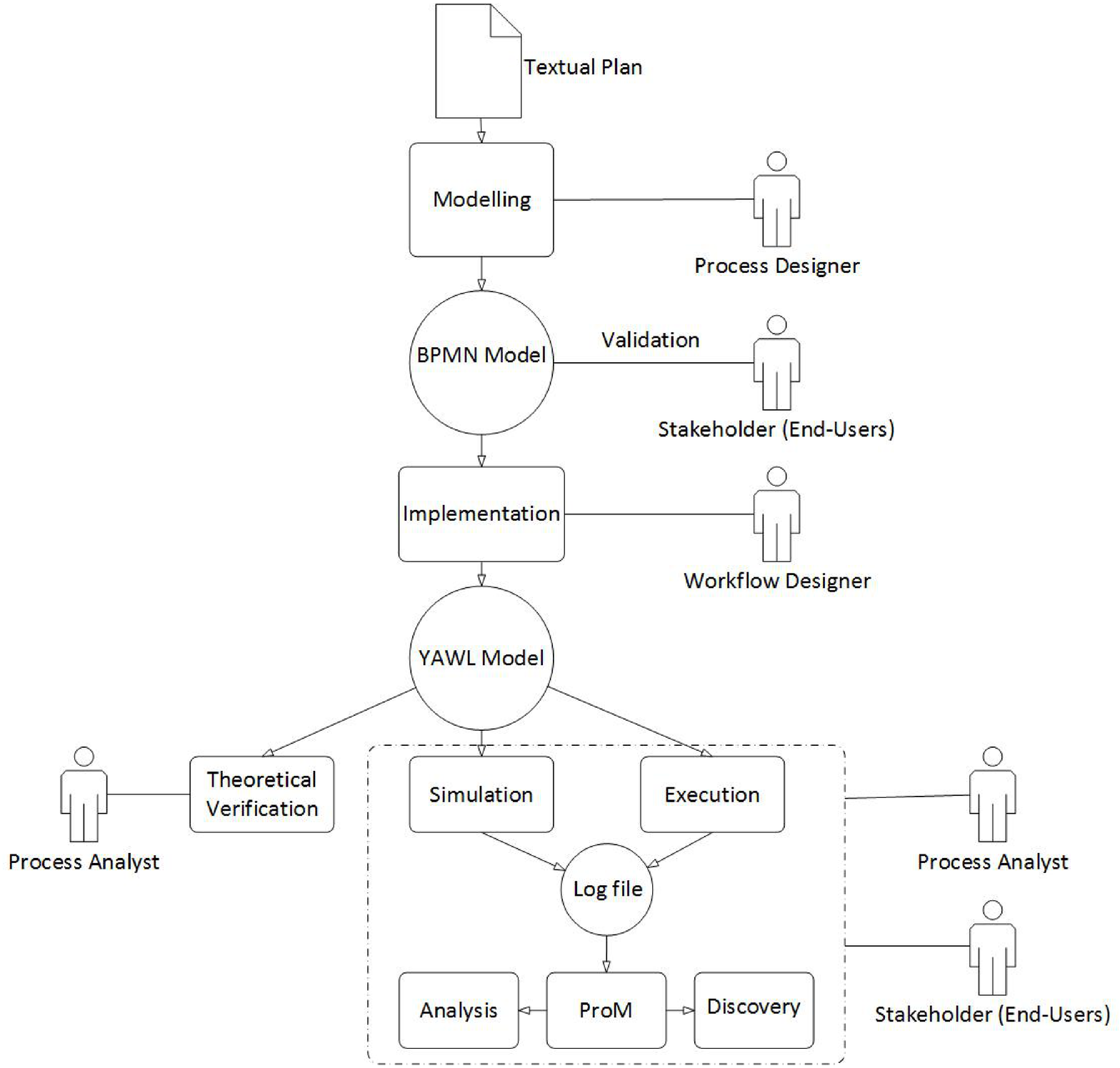}
    \caption{Crisis plan lifecycle}
    \label{LifeCycleProcessingCrisis}
\end{figure}

In this paper, we follow the lifecycle given in Fig.~\ref{LifeCycleProcessingCrisis}. We give a first abstract model for end users thanks to BPMN, and then derived it in YAWL to verify and simulate it. The BPMN model is devoted to end-users (crisis stakeholders) to check the completeness of the plan (actors, tasks) and some coordination issues. However this model does not allow end-users to check the operationnel semantics of it. The transformation of this model in YAWL could help to simulate, analyse and check the dynamic of the plan: task allocation, synchronizatin issues, temporal constraints.

In BPMN, events are represented by circles. \emph{Start Event} is presented by a green circle, while \emph{End Event} is presented by a red circle. An activity (such as T1: “Detect tsunami risk”) is represented with a rounded-corner rectangle and may be performed by an actor (role in fact such as “Institute of Geophysics”). All the activities of a role are included in the same rectangular container called lane. Roles belonging to the same organization are gathered in a big rectangle (called pool) such as the “Ho Chi Minh City” organization that includes “Local administration”, “Local Civil Defense Forces”,... The activities are linked by arrows and gateways (diamond) to represent control structure: $\rightarrow$ represents sequence, + parallelism, and x alternative. 
The dashed arrow with an open arrowhead outline the flow of messages between two tasks. Our result is depicted in Fig. 3. 

When the same task is done by two different actors, we duplicate it instead of creating an abstract actor including both actors. In the Fig. 3, for example, we want to express the task \emph{Evacuate people} realized by two actors (Military and Police) is represented by two tasks T8 and T8' one for each actor.

\begin{figure}[h]
	\centering
    \includegraphics[scale=0.2]{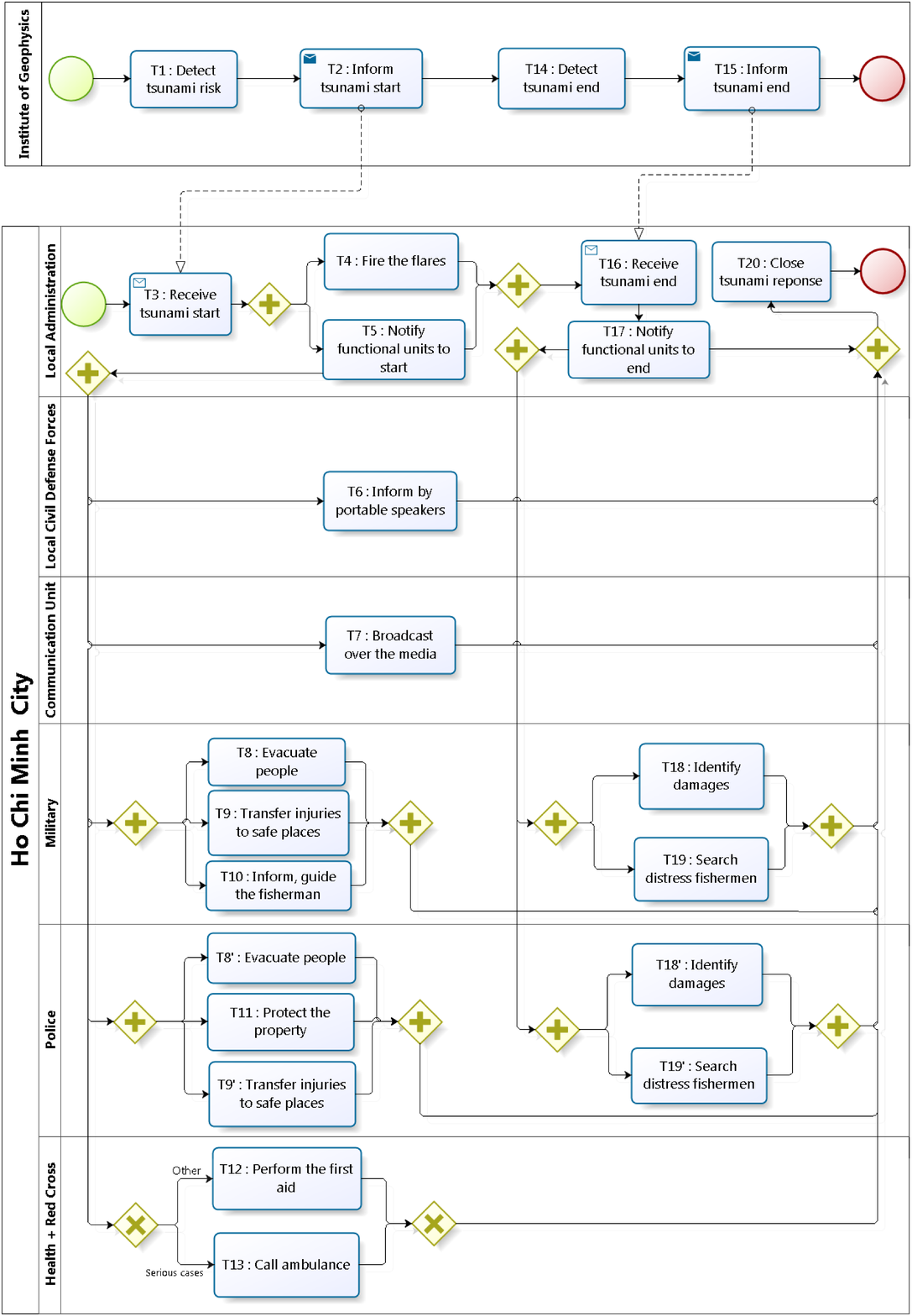}
    \caption{BMPN representation of a Reponse plan to an Ho Chi Minh City Tsunami}
    \label{figurelabel}
\end{figure}

This model obviously eases accountability and awareness. Indeed each actor has a view of his tasks, the links with partners, the situation of extra-works (e.g. \emph{T8}, \emph{T9}, \emph{T10} are tasks performed in parallel by Military and this requires probably more resources and high responsibility).

This model is a shared artefact that could be used for negotiation of resources or improve coordination for similar tasks done by different actors (\emph{T8}, \emph{T8'}).

\section{Engineering Tsunami Response Plans}
In this section, we will show how to simulate predefined plans and how to mine/discover them from event logs.
\subsection{Simulation with YAWL}

From now, we will use YAWL\footnote{\url{http://www.yawlfoundation.org/}} a more operational formalism associated with a Workflow Management System.

YAWL enables the design of Petri-Net based process models on top of which structural analysis are possible. We do not develop this aspect here but we can mention for example that YAWL offers means to check the soundness property which includes three properties: i) process, when started, can always complete, ii) it should not have any other tasks still running for that process when the process ends, iii) the process should not contain tasks that will never be executed (no dead transitions).

The design of YAWL models includes three perspectives: i) the process perspective (tasks and their links), ii) the informational perspective: data used and produced by each task, iii) the organizational perspective defining the roles involved in the process, the actors allowed to play the roles, and the allocation policy of tasks to actors. In this paper, we limit ourselves to the process perspective (see Fig. 4).

YAWL also includes an engine to simulate or execute in a distributed way the processes. The engine automatically allocates tasks to  actors' through their worklists according to the predefined policy, and when the engine is informed of the tasks' evolution, it automatically makes the process evolve. Providing he has a user account, each actor should be connected to his worklist through a web-browser for reading its assigned tasks, accepting or not their realization or not, and notifying of their completion.



\begin{figure}[h]
	\centering
    \includegraphics[scale=0.23]{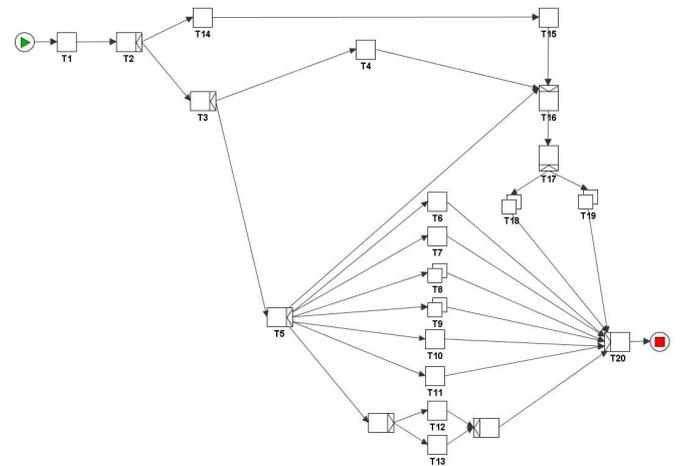}
    \caption{A response Plan for an Ho Chi Minh City Tsunami realized with the YAWL Editor}
   	\label{figurelabel}
\end{figure}   

Fig. 4 describes the tsunami plan according to the YAWL formalism (see Fig. 5 for the meaning of the symbols). Let us remark that YAWL allows to express multiple atomic task which allows to run multiple instances of a task concurrently. This feature is useful to represent for example the evacuate task made both by Military and Police. Parallelism, choice and synchronisation are represented respectively by and-split, or-split and join tasks.

\begin{figure}[h]
	\centering
    \includegraphics[scale=0.4]{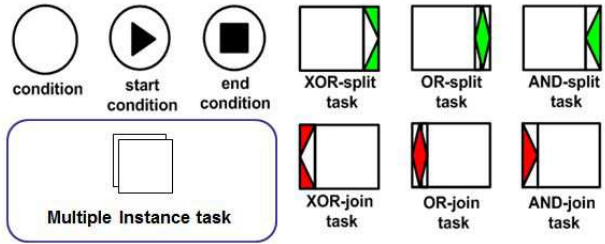}
    \caption{Symbols used in YAWL Editor}
   	\label{figurelabel}
\end{figure} 





Once the three perspectives are filled, executions as well as simulations are possible. A real execution requires that each actor, even in the field, is equipped with a web browser (in mobile phone for example) to interact in a remote mode with the system. A simulation can be performed in a distributed mode or in a centralised way (same space and same computer). During the simulation, we can visualize the state of the process: tasks performed, pending tasks, ... During a simulation or an execution the system records information in event logs that can be exploited for analysing and mining purposes, see next section.

Simulation, when it involves stakeholders could support real-world exercices and improve the preparedness phase, underline coordination and resource issues, and improve the plan.

\subsection{Mining Processes with {\sc ProM}}
Simulation could be linked with the {\sc ProM} environment. {\sc ProM} is an open-source framework implementing process mining tools. Through this environment, it is possible to analyse YAWL processes after their runs (execution or simulation) by examining the facts recorded on the event logs. A process run could be compliant with a predefined model or deviate from it. Also it is possible to translate real world events in a event log to discover process models to be executed or simulated with YAWL. It provides several views on the same process at different abstraction levels. It can discover  organizational and process models. The discovered models could be used for: 1) identifying and understanding new models that have emerged and that may constitute good practices, 2) analysing the compliance between a priori models and the performed ones, 3) improving existing models by analysing performance data (tasks' costs, durations, ...). 

A process may be simulated (or executed) several times and each run gives rise to a case. A case consists of many events such that each event relates precisely to one case. Events within a case are ordered. Events can have attributes. Examples of typical attribute names are: \emph{activity/task}, \emph{timestamp}, \emph{costs} and \emph{resource}. In our work, we use {\sc ProM} version 6.2 that uses the XES and MXML formats for its input event logs.

An abstract view of the event log (without timestamps, actors) is given in Table II.


We assume that the tsunami response system above is operating. During operating time, it generates the event log. For one given case, it does not necessarily perform all the tasks from T1 to T20 because they are alternatives and also because some tasks could be ignored. Here we consider two cases, i.e. two disasters simulations.

\begin{table}[h]
\label{Tasks}
\begin{center}
\begin{tabular}{|c|c|c|c|}
	\hline
	\textbf{CASE} & \textbf{TASK} & \textbf{CASE} & \textbf{TASK}\\
	\hline
	1 & T1 & 2 & T1\\
	\hline
	1 & T2 & 2 & T2\\
	\hline
	1 & T3 & 2 & T3\\
	\hline
	1 & T4 & 2 & T5\\
	\hline
	1 & T5 & 2 & T4\\
	\hline
	1 & T6 & 2 & T7\\
	\hline
	1 & T7 & 2 & T8\\
	\hline
	1 & T8 & 2 & T6\\
	\hline	
	1 & T9 & 2 & T10\\
	\hline	
	1 & T10 & 2 & T9\\
	\hline
	1 & T8' & 2 & T11\\
	\hline
	1 & T11 & 2 & T12\\
	\hline
	1 & T9' & 2 & T14\\
	\hline
	1 & T12 & 2 & T15\\
	\hline
	1 & T13 & 2 & T16\\
	\hline
	1 & T14 & 2 & T17\\
	\hline
	1 & T15 & 2 & T19\\
	\hline
	1 & T16 & 2 & T18'\\
	\hline
	1 & T17 & 2 & T20\\	
	\hline			 
	1 & T18 & - & -\\
	\hline
	1 & T18' & - & -\\
	\hline
	1 & T19 & - & -\\
	\hline
	1 & T19' & - & -\\
	\hline
	1 & T20 & - & -\\
	\hline
\end{tabular}

\end{center}
\caption{Tsunami response process with two cases}
\end{table}

This file is then used as input for {\sc ProM}. We use the plugins for Petri Net and Social Network to analyse this event log. By this way, we rebuild the process from the event log and compare it with the original design in YAWL.

As depicted in Fig. 6, we mine the event log to create the Petri Net using the Alpha-algorithm (\cite{Van:2011:process}). As you can see, the resulting diagram is different from the one designed with YAWL Editor but they are compliant. The reason of the difference between two diagrams is due to the fact that in our event log we only have two cases that doesn't describe all the the possibilites of alternative and parallelism.

From an organizational point of view, {\sc ProM} also discovers sociograms. A sociogram is a graph (P, R) where P is the set of actors and R a relation between actors, in our case it represents the work transfer from an actor A to an actor B. The graph may be weighted to underline the intensity of relationships. In this case, an additional function W is added to assign a value to each element of R. This relationship may also represent different kinds of collaboration: \emph{work transfer, collaboration on  common cases, identical profile among actors}.

\begin{figure}[h]
	\centering
    \includegraphics[scale=0.26]{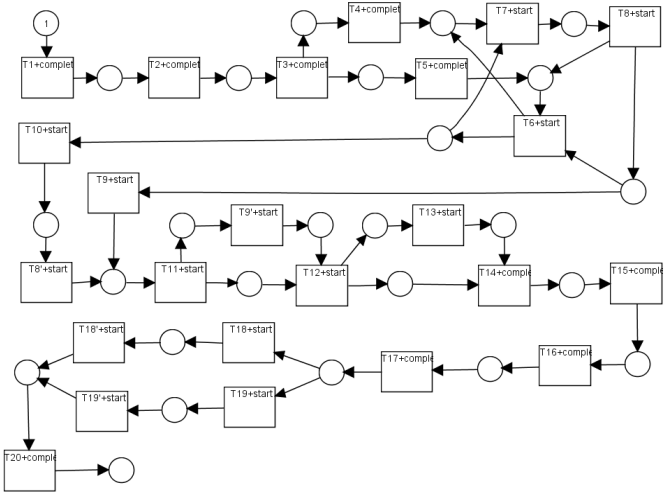}
    \caption{Plan Mining using the Alpha-algorithm plugin of ProM 6.2}
   	\label{figurelabel}
\end{figure}  

In Fig. 7, we present a sociogram built thanks to a specific plug-in called: \emph{Mine for a Handover-of-Work Social Network}. It describes the work transfer relation between our 7 actors. This type of graphs is useful to analyse the structure of the interactions between actors, the level of their involvement in the process and the dependencies of their tasks. We can notice the important involvement of the Military that can be verified by the highest degree of the corresponding node.

\begin{figure}[h]
	\centering
    \includegraphics[scale=0.25]{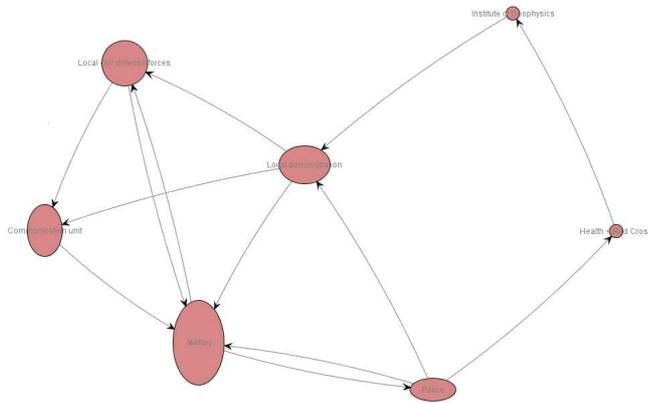}
    \caption{Mine for a Handover-of-Work Social Network plugin of {\sc ProM} 6.2}
   	\label{figurelabel}
\end{figure}   


\section{CONCLUSION}

In this paper, we introduce an approach to model crisis resolution plan from a textual guideline. We follow a process based approach using workflow technology. This technology provides several representations that ease user validation, simulation and analysis. This work has been conducted without involving real crisis stakeholders. We should now try to make more realistic simulations involving the final users and taking into account their feedback.



\addtolength{\textheight}{-12cm}  











\bibliographystyle{ieeetr}
\bibliography{LeNguyenTuanThanh}

\begin{thebibliography}{10}

\bibitem{Ca:2008:tsunami:vietnam}
V.~T. Ca and N.~D. Xuyen, ``Tsunami risk along vietnamese coast,'' {\em Journal
  of Water Resources an Enviromental Engineering}, pp.~24--33, 2008.

\bibitem{Smith:2011:OC:1964041.1964044}
B.~L. Smith, V.~Tamma, and M.~Wooldridge, ``An ontology for coordination,''
  {\em Applied Artificial Intelligence}, vol.~25, pp.~235--265, Mar. 2011.

\bibitem{Van:2003:business:survey}
W.~M. Van Der~Aalst, A.~H. Ter~Hofstede, and M.~Weske, ``Business process
  management: A survey,'' in {\em Business Process Management} (W.~Aalst and
  M.~Weske, eds.), vol.~2678 of {\em Lecture Notes in Computer Science},
  pp.~1--12, Springer Berlin Heidelberg, 2003.

\bibitem{Sell:2009:workflow:emergency}
C.~Sell and I.~Braun, ``Using a workflow management system to manage emergency
  plans,'' in {\em Proceedings of the 6th International ISCRAM Conference},
  vol.~41, (Sweden), May 2009.

\bibitem{Chen:2008:coordination:emergency}
R.~Chen, R.~Sharman, H.~R. Rao, and S.~J. Upadhyaya, ``Coordination in
  emergency response management,'' {\em Communications of the ACM}, vol.~51,
  no.~5, pp.~66--73, 2008.

\bibitem{Hanachi:2012:CTR:2360026.2360677}
C.~Hanachi, F.~Charoy, and S.~Stinckwich, ``Ct2cm track report: 2nd track on
  collaborative technology for coordinating crisis management,'' in {\em
  Proceedings of the 2012 IEEE 21st International Workshop on Enabling
  Technologies: Infrastructure for Collaborative Enterprises}, WETICE '12,
  (Washington, DC, USA), pp.~349--351, IEEE Computer Society, 2012.

\bibitem{Omicini:2001:coordination:internetagents}
A.~Omicini, F.~Zambonelli, M.~Klusch, and R.~Tolksdorf, eds., {\em Coordination
  of Internet Agents: Models, Technologies, and Applications}.
\newblock Springer, 2001.

\bibitem{Malone:1994:coordination:interdisciplinary}
T.~W. Malone and K.~Crowston, ``The interdisciplinary study of coordination,''
  {\em ACM Computing Surveys (CSUR)}, vol.~26, pp.~87--119, March 1994.

\bibitem{Busi:2001:coordination:models}
N.~Busi, P.~Ciancarini, R.~Gorrieri, and G.~Zavattaro, ``Coordination models: A
  guided tour,'' in {\em Coordination of Internet agents}, pp.~6--24,
  Springer-Verlag, 2001.

\bibitem{Franke:2013:coordinatio:framework:dynamic}
J.~Franke, F.~Charoy, and P.~El~Khoury, ``Framework for coordination of
  activities in dynamic situations,'' {\em Enterprise Information Systems -
  Information Systems for Enterprise Integration, Interoperability and
  Networking: Theory and Applications}, vol.~7, pp.~33--60, February 2013.

\bibitem{Schonenberg:2008:taxonomy}
H.~Schonenberg, R.~Mans, N.~Russell, N.~Mulyar, and W.~M. van~der Aalst,
  ``Towards a taxonomy of process flexibility.,'' in {\em CAiSE Forum},
  vol.~344, pp.~81--84, 2008.

\bibitem{Franke:2011:coordination:conflict}
J.~Franke, F.~Charoy, and C.~Ulmer, ``Handling conflicts in autonomous
  coordination of distributed collaborative activities,'' in {\em Enabling
  Technologies: Infrastructure for Collaborative Enterprises (WETICE), 2011
  20th IEEE International Workshops}, pp.~319--326, IEEE, 2011.

\bibitem{Ma:2012:research:management:disaster}
Y.~Ma, F.~Duan, H.~Liu, J.~Tian, and G.~Duan, ``Research on the management and
  expression of disaster data,'' in {\em Geoinformatics (GEOINFORMATICS), 2012
  20th International Conference on}, pp.~1--5, IEEE, 2012.

\bibitem{Catarci:2011:workpad:process:disaster}
T.~Catarci, M.~de~Leoni, A.~Marrella, M.~Mecella, A.~Russo, R.~Steinmann, and
  M.~Bortenschlager, ``Workpad: process management and geo-collaboration help
  disaster response,'' {\em International Journal of Information Systems for
  Crisis Response and Management (IJISCRAM)}, vol.~3, no.~1, pp.~32--49, 2011.

\bibitem{PCHCMCity:2012:decision:3558QDUBND}
P.~C. of~Ho~Chi Minh~City, ``Plan to prevent, respond and overcome the
  consequences of the earthquake, tsunami on the city (vietnamese version),''
  July 2012.

\bibitem{Van:2011:process}
W.~M. Van~der Aalst, M.~de~Leoni, and A.~H. ter Hofstede, ``Process mining and
  visual analytics: Breathing life into business process models,'' {\em BPM
  Center Report BPM-11-15, BPMcenter. org}, 2011.

\end{thebibliography}

\end{document}